\documentclass[aps,showpacs,pra,twocolumn]{revtex4}

\usepackage{epsfig,amssymb,amscd,amsmath}

\usepackage{subfigure}

\usepackage{epsfig}

\usepackage{graphicx}

\newcommand{\Tr}{{\mbox{Tr}}}

\newcommand{\Id}{1\hspace{-0.56ex}{\rm I}}

\newcommand{\be}{\begin{equation}}

\newcommand{\ee}{\end{equation}}

\newcommand{\ket}[1]{ | \, #1  \rangle}

\newcommand{\bra}[1]{ \langle #1 \,  |}

\newcommand{\bea}{\begin{eqnarray} }

\newcommand{\eea}{\end{eqnarray} }

\begin{document}

\bibliographystyle{prsty}

\title{When Non-Gaussian States are Gaussian: Generalization of Non-Separability Criterion for Continuous Variables}

\author{Derek McHugh$^{1}$, Vladim\'\i r Bu\v zek$^{1,2}$, and M\'ario Ziman$^{1,2,3}$ }

\affiliation{ $^1$Research Center for Quantum Information, Slovak
Academy of Sciences, D\'ubravsk\'a cesta 9, 84511 Bratislava,
Slovakia\\
$^2$Quniverse, L\'\i\v s\v cie \'udolie 116, 841 04 Bratislava, Slovakia\\
$^3$Faculty of Informatics, Masaryk University, Botanick\'a 68a,
60200 Brno, Czech Republic}

\received{\today}

\begin{abstract}

We present a class of non-Gaussian two-mode continuous variable
states for which the separability criterion for Gaussian states can
be employed to detect whether they are separable or not. These
states reduce to the two-mode Gaussian states as a special case.

\end{abstract}




\maketitle


Whether a quantum state is entangled or not represents a very
important question in quantum-information theory. Such knowledge
reveals whether one can take advantage of the non-local properties
of the state in quantum protocols such as quantum teleportation
\cite{tele} and quantum cryptography \cite{crypt}. This issue has
been dealt with by many authors in recent years primarily in qubit
systems where the Peres-Horodecki partial transpose separability
condition was the first method to figure out if a two-qubit state
was separable \cite{peres}. In general, for $N$ qubits the solution
is not known. Continuous variable systems have proven to be an
extremely useful setting for quantum cryptography and communication
\cite{vloock}. In these protocols entangled states are required and
the question of separability arises naturally. For two-mode systems
separability criteria for Gaussian states were established in Refs.
\cite{simon,cirac} which proved to be both necessary and sufficient.
More recently necessary and sufficient conditions for the partial transposition
of bipartite harmonic quantum states to be nonnegative are formulated in Ref. \cite{vogel,miran} and
separability criteria based on uncertainty relations for two-mode representations of SU(2) and SU(1,1) algebras in Refs. \cite{agarwal,hillery,nha} can be obtained from from the former result as special cases. These criteria have
particularly targeted uncovering whether non-Gaussian states are
separable or not as previous criteria fail to detect relatively
simple entangled states.

\begin{figure}
 \begin{center}
 \setlength{\unitlength}{1cm}
 \includegraphics[width=6cm]{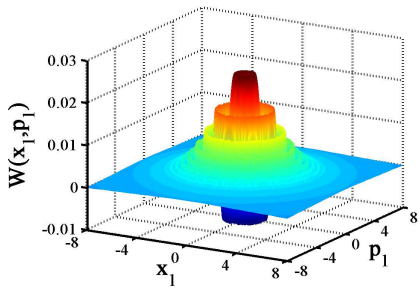}
 \end{center}
 \caption{(Color online) The Wigner function of the reduced state of a $k$-photon two-mode squeezed vacuum state
 as represented in the phase space defined by
 $\hat{x}_1=\hat{a}_1^{\dagger}+\hat{a},\hat{p}_1=i(\hat{a}_1^{\dagger}-\hat{a}_1)$ with $k=3$. This state is non-Gaussian.}
 \label{fig1}
 \end{figure}

In this paper we approach the problem from a different perspective.
The Wigner function of the reduced state of a two-mode quantum state
is shown in Fig. \ref{fig1}. Clearly the Wigner function is
non-Gaussian and hence the two-mode state from which it is derived
is non-Gaussian as well. Say an experimentalist measures (that is reconstructs from
a tomographically complete measurement)
this state in the
laboratory. What exactly can she say about the separability of the
state? At first sight not a great deal. However we will show that in
fact this state's separability is completely known and understood in
terms of its interpretation as a Gaussian state. Furthermore we
introduce a whole class of these states for which the usual two-mode
Gaussian states are a special case.



Gaussian states of a two-mode continuous variable quantum system
have been much studied in the literature and their entanglement
properties are quite well established \cite{seraf,dmmzvb}. Here we
briefly review their characterization. Let
$\hat{a}^{\dagger},\hat{a}$ be the bosonic creation and annihilation
operators acting on a Fock Hilbert space as
$\hat{a}\ket{n}=\sqrt{n}\ket{n-1}$, $\hat{a}^{\dagger}\ket{n}
=\sqrt{n+1}\ket{n+1}$ and satisfying the Weyl-Heisenberg commutation
relations $[\hat{a},\hat{a}^{\dagger}]= \Id$. We can introduce
position and momentum operators $\hat{x}=\hat{x}^{\dagger}+\hat{x}$,
$\hat{p}=i(\hat{a}^{\dagger}-\hat{a})$ (setting $\hbar=1$)
which define phase space
through the continuous range of their eigenvalues. For two modes
these position and momentum operators are defined in each Hilbert
space. We gather these operators into a single
vector $\vec{\hat{x}}=(\hat{x}_1,\hat{p}_1,\hat{x}_2,\hat{p}_2)$ for clarity.
By definition the Wigner function of a Gaussian state, $\rho$, takes
the form $W(x)= \exp[-x'\sigma x'^T/2]/\pi\sqrt{\det\sigma}$, where
$x'=x-\langle \hat{x}\rangle$ and $\sigma_{jk}=(\langle
\hat{x}_j\hat{x}_k+\hat{x}_k\hat{x}_j\rangle)/2-\langle
\hat{x}_j\rangle\langle \hat{x}_k\rangle$ is called the covariance
matrix with $\langle \hat{O}\rangle=\Tr[\rho \hat{O}]$. The Wigner
function thus depends only on the first and second moments of the
position and momentum operators \cite{marcinkiewicz}.
Furthermore, via local operations
the covariance matrix can be brought to the simpler form
\be\sigma_{sf}=\left(\begin{array}{cccc}b_1&0&c_1&0\\
0&b_2&0&c_2\\ c_1&0&d_1&0\\
0&c_2&0&d_2\end{array}\right)\label{CMsf},\ee where $b_i,c_i$ and
$d_i$ satisfy $\frac{b_1-1}{d_1-1}=\frac{b_2-1}{d_2-1}$ and $\mid
c_1\mid- \mid c_2\mid = \sqrt{(b_1-1)(d_1-1)}-\sqrt{(b_2-1)(d_2-1)}$
as shown in Ref. \cite{cirac}. The necessary and sufficient
condition for the state to be separable is then
\be\langle(\Delta
\hat{u})^2\rangle+\langle(\Delta\hat{v})^2\rangle \ge
q_0^2+\frac{1}{q_0^2}\label{sepcrit},
\ee
with $\hat{u}=q_0\hat{x}_1-\frac{c_1}{\mid c_1\mid
q_0}\hat{x}_2$, $\hat{v}=q_0\hat{p}_1-\frac{c_2}{\mid c_2\mid
q_0}\hat{p}_2,$ and $q_0^2=\sqrt{(d_i-1)/(b_i-1)}$. Actually this
expression simplifies to $\sum_{j=1}^2\sqrt{(b_j-1)(d_j-1)}
\ge\sum_{j=1}^2\mid c_j\mid$ when both $b_1-1\ge 0$ and
$d_1-1\ge0$ or $\sum_{j=1}^2(-1)^j\sqrt{(b_j-1)(d_j-1)}
\ge\sum_{j=1}^2\mid c_j\mid$ otherwise. In the following section we impose an additional
criterion: we will relate the degree of entanglement of Gaussian states to an energy of these states.

Gaussian states are an important class of states, but some
quantum protocols require \cite{paris} the usage of non-Gaussian states.
These states are more difficult to
handle mathematically than Gaussian states which only require
knowledge of the first and second moments of a finite set of
observables. One of the reasons why it is more difficult to
analyse non-Gaussian states
is that they are characterized by an infinite set of non-zero cumulants
i.e. higher-order moments of system observables cannot
be expressed in terms of the first and second order moments.
What is important to note is that in order to define
Gaussian states one needs to be able to construct observables from
creation and annihilation operators which satisfy the
Weyl-Heisenberg commutation relations. It is therefore possible to
choose other operators satisfying these commutation relations and
use these to construct new observables whose eigenvalues define a
completely different phase space. In particular we can choose
multi-photon operators like
\be \hat{A}^{(k)\dagger} = \sqrt{\left[\left[
\frac{\hat{N}}{k} \right]\right]\frac{(\hat{N}-k)!}{\hat{N}!}}
\hat{a}^{\dagger k}, \ee as in Ref. \cite{brandt},
satisfying $[\hat{A}^{(k)},\hat{A}^{(k)\dagger}]= \Id$
where $\hat{N}=\hat{a}^{\dagger}\hat{a}$,
$[[\hat{N}/k]]=\sum_n [[n/k]]\ket{n}\bra{n}$,
$(\hat N-k)!/\hat N!=\sum_n \frac{(n-k)!}{n!}\ket{n}\bra{n}$,
and $[[n/k]]$ denotes the largest positive integer less than or equal to $n/k$ . They are
constructed in such a way as to create or annihilate $k$ photons at
a time. These operators can also be interpreted as acting on a
``multi-photon Fock space'', $\tilde{\cal H}$, as
$\hat{A}^{(k)\dagger}\ket{n}_k=\sqrt{n+1}\ket{n+1}_k$,
$\hat{A^{(k)}}\ket{n}_k =\sqrt{n}\ket{n-1}_k$ where the subscript on
the multi-photon number state $\ket{n}_k$ indicates we are referring
to those states satisfying \be\ket{n}_k= \frac{\hat{A}^{(k)\dagger
n}}{\sqrt{n!}} \ket{0}_k.\ee Their action on the usual Fock space is
given by \bea
\nonumber \hat{A}^{(k)\dagger} \ket{nk+m} &=& \sqrt{n+1}\ket{(n+1)k+m}\\
\hat{A}^{(k)}\ket{nk+m}&=& \sqrt{n}\ket{(n-1)k +m},\eea with $n\ge
0$ and $0\le m\le k-1$.  Thus, for the multiphoton number
operators $\hat{N}^{(k)}=\hat{A}^{(k)^\dagger}\hat{A}^{(k)}$
each eigenvalue is k-times degenerated (including the vacua, i.e.
$\hat{A}^{(k)}\ket{m}=0$ for $m<k$).

As before we can construct position and momentum operators
 (acting on $\tilde{\cal H}$ only) for the two modes
$\vec{X}^{(k)}=(\hat{X}_1^{(k)},\hat{P}_1^{(k)},\hat{X}_2^{(k)},
\hat{P}_2^{(k)})$
with $\hat{X}_j^{(k)}=\hat{A}_j^{(k)\dagger}+\hat{A}_j^{(k)}$,
$\hat{P}_j^{(k)}=i(\hat{A}_i^{(k)\dagger}-\hat{A}_j^{(k)})$ and the eigenvalues
of these operators define the phase space. In this phase space we
have Gaussian states whose Wigner functions are Gaussian while if we
represent the states using a Wigner function in the usual phase
space the states are highly non-Gaussian. The separability criterion
from Ref. \cite{cirac} carries over here so that given a state has a covariance matrix in the form \be\sigma_{sf}=\left(\begin{array}{cccc}B^{(k)}_1&0&C^{(k)}_1&0\\
0&B^{(k)}_2&0&C^{(k)}_2\\ C^{(k)}_1&0&D^{(k)}_1&0\\
0&C^{(k)}_2&0&D^{(k)}_2\end{array}\right),\label{CM}\ee
we know it is separable if \be \sum_{j=1}^2\sqrt{(B^{(k)}_j-1)(D^{(k)}_j-1)}\ge
\sum_{j=1}^2\mid C^{(k)}_j\mid,\label{sepm}\ee when both
$B^{(k)}_1-1\ge 0$ and $D^{(k)}_1-1\ge0$ or \be
\sum_{j=1}^2(-1)^j\sqrt{(B^{(k)}_j-1)(D^{(k)}_j-1)}\ge
\sum_{j=1}^2\mid C^{(k)}_j\mid,\ee otherwise. Of course the local operations required
to transform the state under consideration to one which can be
represented by such a covariance matrix are non-Gaussian operations
in terms of the creation and annihilation operators
$\hat{a}_1^{\dagger}, \hat{a}_1,\hat{a}_2^{\dagger},\hat{a}_2$.
Experimentally the operations could be constructed as proposed in
Ref. \cite{braunlloyd}.

\begin{figure}
 \begin{center}
 \setlength{\unitlength}{1cm}
 \includegraphics[width=6cm]{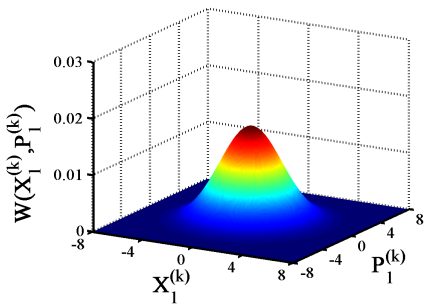}
 \end{center}
 \caption{(Color online) The Wigner function of the reduced state of a $k$-photon two-mode squeezed vacuum state
 as represented in the phase space defined by
 $\hat{X}_1=\hat{A}_1^{(k)\dagger}+\hat{A}^{(k)},\hat{P}_1=i(\hat{A}_1^{(k)\dagger}- 
 \hat{A}_1^{(k)})$ with $k=3$. The state is a
 thermal state and its Gaussian nature is clearly evident.}
 \label{fig2}
 \end{figure}

As an example we can take the multi-photon two-mode squeezed vacuum
state which has the form
$\ket{\Psi}=\sqrt{1-\gamma^2}\sum_n\gamma^n\ket{n}_k\ket{n}_k$. The
reduced state in either mode is given by
$\rho^{(k)}=(1-\gamma^2)\sum_n\gamma^{2n}\ket{n}_k\bra{n}$, a
thermal state. The Wigner function of this state in the $(X^{(k)}_1,P^{(k)}_1)$ reduced phase space
is shown in Fig.
\ref{fig2} and has the expected Gaussian form. However a simple
calculation shows us that \bea\nonumber
\ket{\Psi}&=&\sqrt{1-\gamma^2}\sum_n\gamma^n\frac{\hat{A}_1^{(k)\dagger
n}\hat{A}_2^{(k)\dagger n}}{n!}\ket{0}_k\ket{0}_k\\
&=&\sqrt{1-\gamma^2}\sum_n \gamma^n\ket{kn}\ket{kn}\eea giving a
reduced density matrix $\rho=(1-\gamma^2)\sum_n\gamma^{2n}\ket{kn}
\bra{kn}$. The Wigner function for this state in the $(x_1,p_1)$ reduced phase space is plotted in Fig.
\ref{fig1} showing that it is clearly non-Gaussian and even contains
negative parts.

The covariance matrix of the two-mode multi-photon squeezed state is
as in Eq. (\ref{CM}) with $B^{(k)}_i=D^{(k)}_i=\cosh2r$ and
$C^{(k)}_1=\sinh2r=-C^{(k)}_2$ where $\gamma=\tanh r$. The
separability criterion in Eq. (\ref{sepm}) reads $\exp(-2r)\ge1$ and
the state is entangled for $r>0$. How then does the separability
criterion relate to measurements of observables in the usual Fock
space? For example we need to know what $B^{(k)}_1$ is and to this
end we must measure $\langle \hat{X}^{(k)2}_1\rangle$ and $\langle
\hat{X}^{(k)}_1 \rangle$. For the case $k=2$ we find these expectation
values in terms of the operators $\hat{a}_1,\hat{a}^{\dagger}_1$ are
\bea \langle \hat{X}_1^{(k)\dagger2}\rangle &=& \frac{1}{2}
\left(\left\langle\sqrt{\frac{1}{(\hat{N}_1+1)(\hat{N}_1+3)}}\hat{a}_1^4
\right\rangle\right.\\
&+&\nonumber\left.\left\langle\hat{a}_1^{\dagger 4}
\sqrt{\frac{1}{(\hat{N}_1+1)(\hat{N}_1+3)}} \right\rangle
+2\langle\hat{N}_1\rangle +2\Id\right)\eea and  \be\langle
\hat{X}^{(k)}_1\rangle^2=\frac{1}{2}\left(\left\langle\sqrt{\frac{1}{\hat{N}_1+1}}
\hat{a}_1^2\right\rangle+\left\langle\hat{a}_1^{\dagger
2}\sqrt{\frac{1}{\hat{N}_1+1}}\right\rangle\right)^2\ee with
$\hat{N}_1=\hat{a}^{\dagger}_1\hat{a}_1$. While we readily concede that it is a non-trivial task to experimentally measure these expectation values we are motivated by what one is able to say about the separability of a state when full tomography has been carried out and the result is a state such as that in Fig. 1.

For a given squeezing parameter $\gamma$ the two-mode squeezed vacuum state
$\ket{\Phi}=\sqrt{1-\gamma^2}\sum_n\gamma^n\ket{n,n}$ and its multi-photon equivalent
$\ket{\Psi}=\sqrt{1-\gamma^2}\sum_n\gamma^n\ket{n,n}_k$ will posess the same degree of
entanglement, a fact most easily seen using the the von Neumann entropy of the
reduced state of either mode, $ S=-\Tr\rho_1\ln\rho_1=-\Tr\rho^{(k)}_1\ln\rho^{(k)}_1=S^{(k)}$.
It should be noted however that uncovering the two-mode entanglement present 
in each of the above states requires measurement of different observables.
In what follows we will compare two states of a two-mode system at a fixed energy value \cite{dmmzvb}. In terms of the average energy $\langle E\rangle\equiv\bra{\Phi}\sum_i\hat{N}_i
\ket{\Phi}$ with $\hat{N}_i=\hat{a}^{\dagger}_i\hat{a}_i$ the multi-photon squeezed vacuum state  with the same degree of entanglement has average energy $\langle E^{(k)}\rangle=k\langle E\rangle$.
If we fix the energy of both states at $\langle E\rangle=2\bar{n}$ where $\bar{n}=\bra{\Phi}N_i\ket{\Phi}$ then the reduced state of $\ket{\Phi}$ can be written as  $\rho_1=\sum_np_n\ket{n}\bra{n}$ with $p_n=\bar{n}^n/(1+\bar{n})^{1+n}$ and the reduced state of $\ket{\Psi}$ can be similarly written as $\rho_1^{(k)}=\sum_np_n^{(k)}\ket{n}_k\bra{n}$ with
$p_n^{(k)}=\left(\frac{\bar{n}}{k}\right)^n/(1+\frac{\bar{n}}{k})^{1+n}$.
Thus it is clear that for fixed energy the usual two-mode squeezed vacuum is more entangled than
its multi-photon counterpart. Intuitively this makes sense given that the nature of the
multi-photon state does not allow certain quantum correlations to exist; for instance, upon
measurement of the state in the joint number basis there is zero probability to obtain the result
$\ket{kn+m,kn+m}$ for $m<k$ and $n\in\mathbb{Z}^+$. This result also ties in with the fact that 
among all continuous variable states with a given fixed energy, the maximally entangled states are Gaussian \cite{wolfy}.

\begin{figure}
\begin{center}
\setlength{\unitlength}{1cm}
\includegraphics[width=6cm]{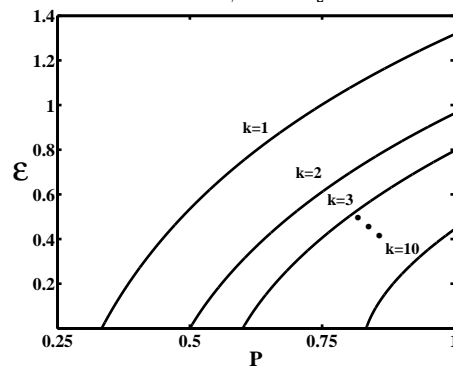}
\end{center}
\caption{The maximally entangled mixed states of the multi-photon Gaussian states plotted at the same average energy $\langle E\rangle=1$ (units are dimensionless) for different values of $k$. The log-negativity, $\mathcal{E}$, is used to measure the entanglement while the purity, $P$, is used to indicate how mixed the state is. }
\label{fig4}
\end{figure}

For mixed states we compare a $k$-photon two-mode mixed state with
a usual two-mode mixed state having the same average energy
$\langle E\rangle$. To get an intuitive sense of the difference
between two such states we look for the minimum purity {(defined
as $P(\omega)={\rm Tr}\omega^2$)} allowed for the $k$-photon mixed
state given this energy $\langle E\rangle$. The dependence on $k$
is
\be
P^{(k)}_{min}=\frac{1}{(\frac{\langle
E\rangle}{k}+1)^2},\ee corresponding to a tensor product of two
multi-photon thermal states.
As $k$ increases the minimum
purity increases asymptotically toward 1 so that the $k$-photon
mixed states tend toward the vacuum in the limit
$k\rightarrow\infty$. To re-enforce this point in Fig. \ref{fig4}
we plot the maximally entangled multi-photon Gaussian mixed states, see Ref. \cite{dmmzvb},
for various values of $k$, all at a fixed mean energy $\langle
E\rangle=1$. Thus we can say that the maximally entangled
$k$-photon mixed states are less entangled than those for $k=1$.
For general mixed states this statement is not always true.


We have presented a large class of non-Gaussian states for which the
existing separability criterion for Gaussian states can be employed
in order to detect their entanglement. In order to clarify our results we recall that
an arbitrary unitary transformation $U:{\cal H}\to{\cal H}$ resulting
in an ``annihilation'' operator $\hat{b}_U=U\hat{a}U^\dagger$ can be exploited
to define another class of non-Gaussian states
${\cal C}_U=\{\varrho:\varrho= U^\dagger\tilde{\varrho}U,\tilde{\varrho}\in
{\cal G}\}$, where $\cal G$ denotes the set of ``standard'' Gaussian states.
Due to the fact that unitary transformations preserve an
operator's spectra and the commutation relations,
the operators $\hat{b}_U,\hat{b}^\dagger_U$ form a representation of the
Weyl-Heisenberg group and Gaussian states with respect to these operators
can be defined. In fact, as it was pointed out
in Ref.~\cite{cirac} the inequality in Eq.(\ref{sepcrit}) provides
a sufficient separability criterion for all operators
$\hat{\mu},\hat{\nu}$ that are locally unitary equivalent to $\hat{u},\hat{v}$,
i.e.  $\hat{\mu} = U\hat{u} U^\dagger$,
$\hat{\nu} = U\hat{v} U^\dagger$, respectively. Moreover, these inequalities
provide the necessary conditions for entanglement for all Gaussian states
defined with respect to new phase-space coordinates, i.e. for all
$\varrho\in{\cal C}_U$.

We have to note that even though the  multi-photon non-Gaussian states analyzed in our paper
seem to be of the similar form as discussed above, there
is a significant difference. The operators $\hat{a}$ and $\hat{A}_k$
are not mutually related by a unitary transformation in the above
sense (for more details see Ref.~\cite{luis93}). In fact, the number operators $\hat{N}_k$ and $\hat{n}$
have different spectra ($\hat{N}_k$ is degenerated). The construction
in our case is based on the fact that the (semi)infinite Hilbert space of the original harmonic 
oscillator can be expressed
as a finite direct sum of (semi)infinite Hilbert spaces that are isomorphic to the
original one, i.e. ${\cal H}=\tilde{\cal H}_0\oplus\cdots\oplus
\tilde{\cal H}_{k-1}$. Here $\tilde{\cal H}_j$ is  a linear span of
vectors ${\cal H}\ni\ket{nk+j}\equiv\ket{n}_{j,k}\in\tilde{\cal H}_j$
($n=0,\dots,\infty$). Physically this means that
we are restricted to states belonging to the subspace
spanned on photon number states separated by a fixed energy $k\hbar\omega$
(representing the energy of $k$ photons). The vacuum for
$\tilde{\cal H}_j$ is represented by the
state $\ket{0+j}=\ket{j}\in{\cal H}$
The linear spaces ${\cal H}$ and $\tilde{\cal H}_j$
are related by a {\em non-bijective} transformation. However, since $\tilde{H}_j$
and $\tilde{\cal H}_{j^\prime}$ are in one-to-one correspondence,
we can write ${\cal H}=\bigoplus_{j=0}^{k-1}\tilde{\cal H}_j
=\tilde{\cal H}_k\otimes{\cal V}_k$ (${\rm dim}\tilde{\cal H}_k=\infty$
and ${\rm dim}{\cal V}_k=k$). Using this notation
the multiphoton annihilation operators are unitarily related
to the original annihilation operator (acting on $\tilde{\cal H}_k$)
via the unitary transformation $\tilde{U}=\sum_{n,m}
(\ket{n}_k\otimes\ket{m})\bra{kn+m}$ \cite{luis93}
performing the transformation from ${\cal H}$ to $\tilde{\cal H}_k\otimes{\cal V}_k$.
In particular, $\tilde{U} \hat{A}^{(k)} \tilde{U}^\dagger =
\tilde{a}\otimes I$.

The Gaussian states are naturally a special case of these non-Gaussian
states as one would expect. For two modes the operation moving from the basis
in which the states have a Gaussian Wigner representation to that
in which they don't is local unitary and as such preserves the entanglement.
This holds for $\hat{b}_U$, but also for multiphoton operators
$\hat{A}^{(k)}$, hence the criterion derived for standard Gaussian states
can be directly applied to multiphoton Gaussian states as it was demonstrated
in the present work. A question remains is how to efficiently verify whether a given
state belongs to a certain sector of the Hilbert space $\tilde{\cal H}_j$ for
a given $k$, or not. The answer can be given by analyzing the expression for the
state under consideration in the Fock basis. If the populated
(i.e. non-vanishing) levels are separated by the same energy
(equivalently, by the same number of photons), then the state belongs
to a multiphoton sector $\tilde{\cal H}$ of the Hilbert space $\cal H$
and its multiphoton Wigner function can be further analyzed.

{\bf Acknowledgement}\\
This work was supported in part by the
European Union projects INTAS-04-77-7289, CONQUEST and QAP,  by
the Slovak Academy of Sciences via the project CE-PI/2/2005, by the
project APVT-99-012304.


\begin{thebibliography}{99}
\bibitem{tele} C. H. Bennett {\it et al.}, 
Phys. Rev. Lett. {\bf 70}, 1895 (1993);
S. L. Braunstein {\it et al.},  
Phys. Rev. Lett. {\bf 80}, 869 (1998).
\bibitem{crypt} C. H. Bennett {\it et al.},
{\it Proc. of IEEE International Conference on Computers, Systems and Signal
Processing, Bangalore, India} (IEEE, New York, 1984), p. 175; M.
Hillery, Phys. Rev. A {\bf 61}, 022309 (2000).
\bibitem{peres} A. Peres, Phys. Rev. Lett. {\bf 77}, 1413 (1996).
\bibitem{vloock} P. van Loock, Fort. der Phys. {\bf 50}, 1177
(2002).
\bibitem{simon} R. Simon, Phys. Rev. Lett. {\bf 84}, 2726 (2001).
\bibitem{cirac} L. M. Duan {\it et al.}, 
Phys. Rev. Lett. {\bf 84}, 2722 (2001).
\bibitem{vogel} E. Schukin {\it et al.}, Phys. Rev. Lett. {\bf 95}, 230502 (2005); E. Schukin {\it et al.}, Phys. Rev. A {\bf 74}, 030302(R) (2006).
\bibitem{miran} A. Miranowicz {\it et al.}, quant-ph/0605001.
\bibitem{agarwal} G. S. Agarwal and A. Biswas, New J. Phys. {\bf 7},
211 (2005).
\bibitem{hillery} M. Hillery {\it et al.},
Phys. Rev. Lett. {\bf 96}, 050503 (2006).
\bibitem{nha} H. Nha and J. Kim, Phys. Rev. A {\bf 74}, 012317
(2006).
\bibitem{seraf} G. Adesso {\it et al.}, 
Phys. Rev. Lett. {\bf 92}, 087901 (2004); G. Adesso {\it et al.}, 
Phys. Rev. A {\bf 70}, 022318 (2004).
\bibitem{dmmzvb} D. McHugh {\it et al.}, 
Phys. Rev. A {\bf 74}, 042303 (2006).
\bibitem{marcinkiewicz} J.Marcinkiewicz, Math. Z. {\bf 44}, 612 (1939).
\bibitem{paris} S. Olivares {\it et al.},
Phys. Rev. A {\bf 70}, 032112 (2004).
\bibitem{brandt} R. A. Brandt {\it et al.},
J. Math. Phys. {\bf 10}, 1168 (1969).
\bibitem{braunlloyd} S. Lloyd {\it et al.},
Phys. Rev. Lett. {\bf 82}, 1784 (1999).
\bibitem{wolfy} M. M. Wolf {\it et al.}, 
Phys. Rev. Lett. {\bf 96}, 080802 (2006).
\bibitem{luis93} A. Luis {\it et al.},
Quantum.Opt. {\bf 8}, 33-41 (1993)
\end{thebibliography}
\end{document}